\documentclass[aps,prl,twocolumn,showpacs,amsmath,amssymb]{revtex4}
\usepackage{dcolumn}
\usepackage{bm}
\usepackage{graphicx}
\usepackage{times}
\usepackage{color}
\usepackage{epstopdf}
\begin{document}

\renewcommand{\section}[1]{{\par\it #1.---}\ignorespaces}
\newcommand{\bfk}{{\bf k}}

\title{Spontaneous gap generation on the surface of weakly interacting topological insulators using nonmagnetic impurities}
\author{Annica M. Black-Schaffer}
\author{Dmitry Yudin}
 \affiliation{Department of Physics and Astronomy, Uppsala University, Box 516, S-751 20 Uppsala, Sweden}
\date{\today}

\begin{abstract}
Strong nonmagnetic impurities on the surface of three-dimensional topological insulators (TIs) generate localized resonance peaks close to the Dirac point. We show that this results in a strongly reduced critical Coulomb interaction strength to reach a magnetic surface state, following a Stoner-like criterion. Thus even weakly interacting TIs host a finite (local) magnetization around strong nonmagnetic impurities. The local magnetization gives rise to a global energy gap, linearly dependent on the maximum value of the magnetization but decreasing with reduced impurity concentration.
\end{abstract}
\maketitle

%
Topological insulators (TIs) are insulating in the bulk but have conducting surfaces due to a nontrivial topology of the bulk band structure \cite{Hasan10, Qi11RMP}.
The surface states are described by a two-dimensional (2D) massless Dirac Hamiltonian \cite{Fu07, Moore_PRB07, Qi08, Hsieh09}, with the momentum locked to the electron spin, and TIs thus belong to the newly emergent class of Dirac Materials \cite{DMreview14}. 
In strong TIs there are only one (or odd number) Dirac cone per surface and the spectrum can only be gapped by impurities or other perturbations breaking time-reversal symmetry \cite{Fu07, Moore_PRB07, Qi08}. Accordingly, magnetic impurities have been predicted to generate an energy gap \cite{Liu09TImagimp}, whereas their nonmagnetic counterparts should neither gap the spectrum nor allow full $180^\circ$ back-scattering, which requires a spin-flip \cite{Roushan09}. 

Despite this clear expected distinction between magnetic and nonmagnetic impurities, no experimental consensus has yet appeared as to the properties of the TI surface state in the presence of impurities. 
Magnetic impurities in the bulk and thin films have been shown to generate features resembling a gap in the surface spectrum \cite{Chen10, Wray11, Xu12}, but no energy gap has been reported for magnetic impurities deposited directly on the surface \cite{Scholz12, Valla12, Honolka12PRL, Schlenk13PRL}. Interestingly, several studies have also reported no significant difference in the behavior of magnetic and nonmagnetic surface impurities \cite{Bianchi11, Valla12}.

Beyond possibly distorting the Dirac surface spectrum by opening an energy gap, impurities have also been shown to induce resonances in the energy spectrum \cite{Biswas10, Black-Schaffer12TIimp, Black-Schaffer12TIimp2}, confirmed experimentally for both nonmagnetic impurities \cite{Teague12, Alpichshev_PRL12} and step edges \cite{Alpichshev11}. Even for a nonmagnetic impurity the resonance peak approaches the Dirac point in the strong scattering limit, where it splits the original Dirac point into two points which move off-center with the resonance peak in-between \cite{Black-Schaffer12TIimp}. 

The resonance peak resulting from a strong nonmagnetic impurity provides a very large density of states (DOS) at the Fermi level in pristine TIs. It therefore seems natural to ask the question if this system is unstable towards spontaneous spin-polarization? The unperturbed Dirac spectrum has a vanishing DOS at the Fermi level and should thus be very stable against a phase transition to a magnetic state. However, in the presence of a low-energy impurity-induced resonance peak, spontaneously breaking time-reversal symmetry and generating a (local) magnetization might lower the energy even for weak electron-electron interactions.

In this Letter we show that in the presence of even weak electron-electron interactions nonmagnetic impurities can generate a magnetic state locally around the impurities. More specifically, the critical interaction strength to reach a spin-polarized state is dependent on the impurity strength and concentration, following a Stoner-like criterion and approaching zero for dilute concentrations of strong impurities. Moreover, we find that the magnetic state induces an energy gap, which is directly proportional to the maximum value of the magnetization, but reduced with decreasing impurity concentration. Thus, nonmagnetic impurities can in the presence of even weak electron-electron interactions spontaneously generate a finite mass in the Dirac surface state of a TI.

For the main calculations we employ a simple and often used tight-binding model for a strong TI consisting of $s$-orbitals on the 3D diamond lattice with nearest neighbor hopping $t$ and spin-orbit coupling $\lambda$ \cite{Fu07}:
%
\begin{align}
\label{eq:H0}
H_0 = \! \! \! \sum_{\langle i,j\rangle,\sigma} \! \! (t+\delta t_{ij}) c^\dagger_{i\sigma}c_{j\sigma} +
\frac{4i\lambda}{a^2} \! \! \! \! \! \sum_{\langle \langle i,j\rangle \rangle,\sigma\sigma'} \! \! \! \! \! c^\dagger_{i\sigma} {\bf s \cdot \! (d}^1_{ij}\times {\bf d}^2_{ij}) c_{j\sigma'}.
\end{align}
Here $c_{i\sigma}^\dagger$ is the creation operator on site $i$ with spin-index $\sigma$, $\sqrt{2}a$ is the cubic cell size with $a=1$ the unit of length, ${\bf s}$ denote the Pauli spin matrices, and ${\bf d}_{ij}^{1,2}$ are the two bond vectors connecting next-nearest neighbor sites $i$ and $j$. We further set $\lambda = 0.3t$ and assume an undoped system. By choosing $\delta t_{ij} = 0.25t$ for only one of the nearest neighbor directions not parallel to (111), a strong TI with a single Dirac surface cone is created \cite{Fu07}. We construct a TI surface by creating a slab in the (111) direction with  ABBCC ... AABBC stacking termination. To avoid cross-talk between the two surfaces we use $\gtrsim 5$ lateral unit cells \cite{Black-Schaffer12TIimp}. 
We furthermore set $t = 2$, which gives the slope $\hbar v_F\approxeq 1$ of the surface Dirac cone.

We study nonmagnetic (potential) impurities on the surface of the TI by creating a rectangular-shaped surface supercell with $n$ sites along each direction, resulting in the surface area $\sqrt{3}n^2a^2/2$. A single nonmagnetic impurity with strength $V$ is then added to the supercell through $H_{\rm imp} = V \sum_\sigma c_{1\sigma}^\dagger c_{1\sigma}$.
We finally incorporate the effect of electron-electron interactions through a Hubbard-$U$ repulsion: $H_U = U\sum_i c_{i\uparrow}^\dagger c_{i\uparrow}c_{i\downarrow}^\dagger c_{i\downarrow}$. This term takes into account only on-site repulsion, but the $1/r$ Coulomb tail has been found to be marginally irrelevant in a renormalization group sense at the interacting fixed point for short-range repulsion \cite{Herbut06,Herbut09}, justifying the approach. 
Below we also show that a complementary continuum model calculation using long-range Coulomb interaction gives qualitatively the same results.
 
For contact interactions purely out-of-plane magnetization has been found to be favorable over having finite in-plane magnetization components \cite{Baum12}. We therefore perform a mean-field decomposition of $H_U$ with only the $z$-component of the magnetization: $m_i = \frac{1}{2}\langle c_{i \uparrow}^\dagger c_{i \uparrow} - c_{i \downarrow}^\dagger c_{i \downarrow}\rangle$ as the order parameter
  \footnote{We have checked that also self-consistently calculating the total charge per site $N_i = \langle c_{i\uparrow}^\dagger c_{i\uparrow} + c_{i\downarrow}^\dagger c_{i\downarrow}\rangle$ does not change the results.}.  
We then solve the full problem with an impurity and electron interactions: $H= H_0 + H_U + H_{\rm imp}$ self-consistently for the magnetization $m_i$ in the whole supercell. In terms of the numerical details we have confirmed convergence with respect to the $k$-point resolution and have found that a Gaussian broadening of $\sigma = 0.005$ gives good resolution for the local density of states (LDOS).
 Due to the low DOS in the pristine TI surface Dirac cone short-range interactions are perturbatively irrelevant and a finite critical coupling is needed to enter a magnetic state \cite{Gonzalez01,Herbut06, Baum12}. We find that the critical interaction strength needed to achieve a finite (surface) magnetization is $U_c \gtrsim 5t$ when no impurities are present. Below we show that the critical interaction strength is significantly reduced, even down to zero, for strong nonmagnetic impurities.

%
\section{Impurity-induced magnetization}
\begin{figure}[htb]
\includegraphics[scale = 1.02]{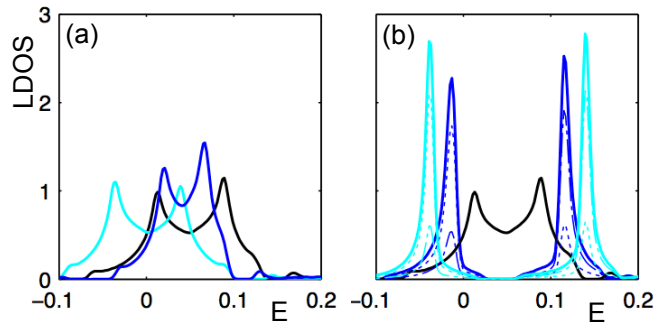}
\caption{\label{fig:LDOS} (Color online). Low-energy LDOS (states/energy/area) averaged over in-plane nearest neighbor sites to a TI surface potential impurity. (a) Non-interacting system with a vacancy and supercell size $n = 10$ (black), $n = 14$ (blue), and a $V = 65t$ impurity and size $n = 10$ (cyan). (b) System with a vacancy and size $n = 10$ for no interactions $U = 0$ (black), $U = 2t$ (blue), and $U = 2.4t$ (cyan), where $U_c = 1.8t$. For the spin-polarized systems the spin-up LDOS (dotted line) and spin-down LDOS (dash-dotted line) are plotted. The linearly dispersing TI surface state LDOS is  $\sim0.01$ for $E =\pm 0.5$ and is not visible. 
}
\end{figure}

Nonmagnetic impurities on the surface of a TI have been shown to induce localized impurity resonances, with the impurity resonance energy scaling as $E_{\rm res} \simeq -1/V$ \cite{Biswas10, Black-Schaffer12TIimp, Teague12, Alpichshev_PRL12}. Thus a vacancy, where $V$ approaches infinity, gives a resonance at the Dirac point, as shown in Fig.~\ref{fig:LDOS}(a) for the non-interacting case. 
Finite impurity concentrations lead to a double-peak resonance, which narrows with decreasing concentration (cp.~black and blue).
It also shifts the resonance slightly past the original Dirac point, since potential impurities induce a small residual overall doping into the system \cite{Black-Schaffer12TIimp}. A resonance peak firmly located at the Dirac point for finite supercells thus requires a large but finite $V$ (cyan).

%

The large DOS around the Fermi level close to a strong potential impurity fundamentally changes the sensitivity to interaction effects. In Fig.~\ref{fig:LDOS}(b) we show how the resonance peak changes between the non-interacting case (black) and $U>U_c$ (blue, cyan). The resonance peak splits into two spin-polarized peaks with an energy gap developing in-between. The peak splitting increases with interaction strength and the spin-polarization is large but not complete. 
The resulting magnetization is strongly localized around the impurity. We find that the magnetization is essentially zero beyond the fourth in-plane neighbors and a surface effect, dying out within six atomic layers (one lateral unit cell). The magnetization is antiferromagnetically aligned between (111) planes but ferromagnetic in each plane, apart form the surface plane where it oscillates with the distance to the impurity.

The critical interaction strength $U_c$ for finite magnetization depends on both impurity strength and concentration, as we will discuss below, but is always significantly reduced from the clean limit for strong impurities, e.g.~in Fig.~\ref{fig:LDOS}(b) $U_c = 1.8t$ compared to $U_c \gtrsim 5t$ without impurities. 
However, a TI can in general not be modeled on a pure bipartite lattice and thus Lieb's theorem \cite{Lieb89}, which states that a site imbalance between the two sublattices gives a finite magnetic moment for any finite Hubbard $U$, does not apply. Thus vacancies can still require a finite $U_c$ to reach a magnetic state in a TI. This is in sharp contrast to idealized graphene, which also have a Dirac dispersion, where a vacancy always generates a finite magnetic moment \cite{Yazyev07, Yazyev10, Ugeda10}. The bipartite limit is reached in Eq.~(\ref{eq:H0}) by setting $\lambda =0$ and we then find $U_c =0$ for a vacancy, but this is not a topological state.

%
\section{Stoner-like instability}
\begin{figure}[htb]
\includegraphics[scale = 1.04]{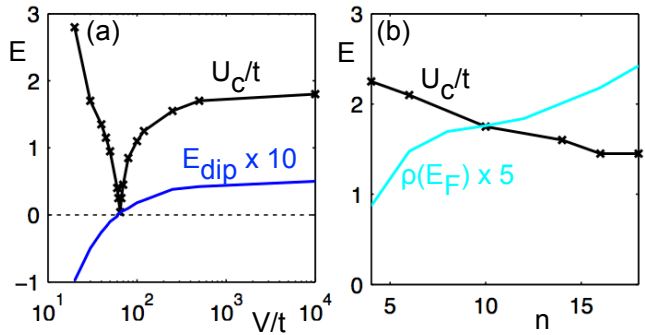}
\caption{\label{fig:Uc} (Color online). Critical interaction strength $U_c/t$ (black, crosses) as function of impurity strength $V/t$ for supercell size $n = 10$ (a) and as function of supercell size $n$ for a vacancy (b). In (a) the energy of the dip (scaled by a factor of 10) in the double-peak resonance is also plotted (blue) along with zero energy (dotted line). In (b) the DOS at the Fermi level $\rho(E_F)$ averaged over in-plane nearest neighbor sites to the impurity (scaled by a factor of 5) is also plotted (cyan).
}
\end{figure}

In order to determine when a finite magnetization is induced by nonmagnetic surface impurities, we plot $U_c$ as function of impurity strength for a fixed impurity concentration in Fig.~\ref{fig:Uc}(a). 
Even though Lieb's theorem does not guarantee $U_c = 0$ for vacancies, we still find that a particular large and finite impurity strength gives essentially $U_c = 0$, and thus finite magnetization even for infinitesimally weakly interacting TIs. 
%
This result can be explained by studying the impurity resonance peak positions in Fig.~\ref{fig:LDOS}(a).
Strong impurities push the resonance peak towards the Fermi level and thus $U_c$ goes down sharply as the resonance state starts to generate a large DOS around the Fermi level. However, a finite concentration of strong impurities also induces a finite residual doping in the system. This leads to the impurity resonance eventually moving past the Fermi level, which reduces the DOS at the Fermi level, and thus $U_c$ increases again when approaching the unitary scattering limit. 
With decreasing impurity concentration the residual doping decreases and the dip in $U_c$ is found at ever higher impurity strengths, such that in the limit of an isolated vacancy $U_c \approx 0$.
To corroborate this picture we also plot the energy of the central dip in the double-peak resonance structure (blue). Clearly, having the resonance peak exactly positioned at the Fermi level, i.e.~the dip at $E = 0$, is extremely well correlated with a vanishing $U_c$. 

In Fig.~\ref{fig:Uc}(b) we investigate more closely the concentration dependence of $U_c$ for vacancies. Decreasing concentration leads to both narrower resonance peaks and smaller residual doping. 
This results in a higher DOS at the Fermi level for vacancies (cyan) and we see how this transfers into $U_c$ steadily decreasing with the supercell size $n$. 
This inverse correlation between DOS at the Fermi level and $U_c$ is a very characteristic feature of Stoner magnetism. The Stoner criterion for a bulk ferromagnetic state reads $U_c \rho(E_F)/2 = 1$, where $\rho(E_F)/2$ is the bulk DOS at the Fermi level $E_F$ for one spin species in the paramagnetic state. 
However, in non-homogenous systems with impurities, impurity-induced Stoner magnetism has been shown to not be sensitive to the precise DOS at $E_F$, but to the whole impurity band if it is narrow enough \cite{Edwards06, Coey10}. We clearly see such an effect in Fig.~\ref{fig:Uc}(a), where $U_c$ is primarily determined by the center of the resonance peak, i.e.~the dip between the two peaks.
We find an approximately constant relation between $U_c$ and $\rho(E_F)$ for a range of different impurity concentrations, but the critical interaction strength is noticeably reduced compared to the bulk Stoner criterion.
Impurity-induced Stoner magnetism has been proposed to generate magnetism for interaction strengths significantly below the bulk Stoner criterion in a number of systems, ranging from hydrogenated graphene \cite{Yazyev07} and C$_{60}$H$_n$ \cite{LeeLee11} to CaB$_6$ \cite{Edwards06}. For example, charge transfer between narrow impurity bands and the bulk has been shown to enhance the propensity for impurity-driven magnetism \cite{Coey10}.
We thus conclude that the spontaneous magnetization in TIs with nonmagnetic impurities can be understood as a Stoner-like instability, with the overall DOS of the resonance peak close to the Fermi level determining the critical interaction strength.

%
\section{Spontaneous mass generation}
The finite magnetization appearing around potential impurities as soon as $U>U_c$ leads to broken time-reversal symmetry. Thus the topological protection of the TI surface state is lost and a mass term, giving rise to an energy gap, is allowed in the TI surface Hamiltonian. 
Intrinsically magnetic impurities placed on a TI surface have been shown to induce a local energy gap \cite{Liu09TImagimp}, and it is reasonable to expect a finite energy gap also for potential impurities when $U>U_c$.

\begin{figure}[htb]
\includegraphics[scale = 0.95]{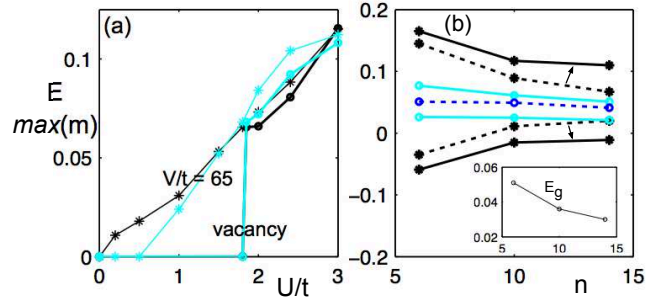}
\caption{\label{fig:Eg} (Color online). (a) Energy gap (cyan) and maximum site magnetization (black) tracking each other essentially perfectly as function of interaction strength $U/t$ for a supercell with $n = 10$ and a vacancy (thick lines, circles) and a $V/t = 65$ impurity (think lines, stars). (The similarity in $y$-values is a coincidence.)
(b) Energies for the two impurity-induced resonance peaks (black, stars) and the in-between dip (blue, circles) for a vacancy in a non-interacting system (dotted lines) and for the corresponding spin-polarized system when $U/U_c = 1.14$ (solid lines) with energy gap edges (cyan, circles) as a function of system size $n$. Small arrows mark the shift of the peak energies appearing at finite spin-polarization. Inset shows the extracted energy gap (difference between cyan lines).
}
\end{figure}

The energy gap $E_g$ could, in principle, depend on the supercell size $n$, the interaction strength $U$, and the impurity strength $V$. Above we have shown that $V$ determines the DOS at the Fermi level which in turn sets $U_c$ through an impurity-induced Stoner mechanism, so we can replace the dependence on $V$ with $U_c$. 
In Fig.~\ref{fig:Eg}(a) we then plot the energy gap $E_g$ (cyan) as function of $U$ for fixed $n$ but two different impurities. We also plot the maximum magnetization ${\rm max}(m)$ (black), which we find to be only a function of $(U/U_c-1)$.
%
For both a vacancy, where $U_c = 1.8t$, and for a $V = 65t$ impurity, where $U_c \approx 0$, the energy gap tracks the maximum magnetization extremely well over the whole range of interaction strengths. The maximum magnetization value is found on the nearest neighbor sites to the impurity (second surface layer), but instead using the average magnetization gives a similarly strong linear dependence between energy gap and magnetization. We thus conclude that the energy gap only depends on the supercell size $n$ and the magnetization, such that $E_g = C(n){\rm max}(m)$, with $C$ being a function of the impurity concentration.

For finite impurity concentrations we find that the energy gap is a global property of the system, i.e.~it does not vary with distance from the impurity. We thus expect the energy gap to decrease when the impurity concentration is decreasing, since the same maximum magnetization will then be responsible for producing an energy gap in a larger area. This is verified in the inset in Fig.~\ref{fig:Eg}(b), where $U/U_c$, or equivalently the maximum magnetization, is kept fixed while the concentration is varied. Thus $C(n)$ decreases for increasing $n$. 
In Fig.~\ref{fig:Eg}(b) we show more details on how the resonance peak structure evolves with impurity concentration for a non-interacting system (dashed) and for a fixed $U/U_c >1$ (solid). The energy gap (cyan) develops around the initial dip (blue) in the double-peak resonance. 
In order to accommodate the finite energy gap the two resonance peaks (black) are pushed out to larger energies. With decreasing impurity concentration the peak-peak distance is decreased along with the overall energy gap.

We can now also draw qualitative conclusions in the limit of isolated impurities. The overall residual doping is then diminished and thus only vacancies produce impurity-induced resonance peaks close to zero energy. Such resonance peaks are very sharp so $U_c$ approaches zero for isolated vacancies. Thus even extremely weakly interacting TIs will have a finite magnetization around isolated vacancies. However, even though ${\rm max}(m)$ only depends on $U/U_c$ and can therefore be large even for very weakly interacting TIs, the size of the induced energy gap will be severely limited by the small function $C(n)$. We thus do not expect any sizable energy gap in the limit of isolated vacancies. 

%
\section{Continuum model}
To complement the above supercell calculations using a Hubbard-$U$ interaction, we also study the effect of long-range Coulomb interactions in a continuum model while treating the impurities within the coherent potential approximation (CPA). The kinetic part is here described by a  2D gapless Dirac term, whereas the electron-electron interaction strength is characterized by the dimensionless coupling $g=e^2/(\epsilon_0 \hbar v_F)$, where $e$ is the charge of the electron and $\epsilon_0$ the dielectric constant of the system. We further here assume that the impurities are randomly distributed over the lattice sites, with the on-site potential taking values $V$ and 0 with probabilities $c$ and $1-c$, respectively, which allows for a binary alloy analogy with the impurity concentration directly linked to $c$. 

For a strongly scattering medium with low impurity concentration the effect of the impurities can be incorporated in a self-consistent manner using CPA, which is based on a single-site approximation in a multiple scattering description. It takes into account terms linear in $c$ but disregards scattering off impurity clusters. 
Using the self-energy $\sigma_{\rm CPA}(\varepsilon)$, which is assumed to be translationally invariant after configurational averaging, as well as spin-independent and site-diagonal, the single particle Green's function of the disordered (but non-interacting) system can be written as  $G_{\rm CPA}^\pm(\varepsilon,{\bm k})=\left(\varepsilon-\sigma_{\rm CPA}(\varepsilon)-\varepsilon_{\bm k}^\pm\right)^{-1}$, where $\varepsilon_{\bm k}^\pm=\pm |\bm{k}|$ is the bare dispersion relation and $\sigma_{\rm CPA}(\varepsilon)$ is determined self-consistently \footnote{See supplementary material for more information.}. 
For small concentrations of vacancies (${\rm Im}\,\sigma\ll k_c$ with a cut-off $k_c$ for integral regularization) we obtain a finite concentration of charge carriers in the vicinity of $\varepsilon=0$. Calculating the change in energy relative to the paramagnetic state we find ferromagnetism when $g_{c}\sim3.7$, to be compared to $g_c\sim 5$ in the absence of disorder. Thus, also a continuum model with long-range Coulomb interactions facilitates a finite magnetization for noticeable weaker electron-electron interactions in the presence of impurities treated within the CPA. Note that this is in spite of the CPA averaging over impurity configurations, thus not explicitly relying on localized impurity-induced resonance states.

In summary we have shown that strong nonmagnetic impurities on the surface of a TI can induce a finite magnetization and energy gap in the presence of even weak electron-electron interactions. Strong impurities and also vacancies give rise to localized resonance peaks around the Dirac point. The resulting increased low-energy DOS leads to a strongly reduced critical interaction strength to reach a magnetic surface state. Thus even very weakly interacting TIs will have a finite magnetization emerging around strong nonmagnetic impurities. 
The finite magnetization gives rise to a global energy gap which is linearly dependent on the maximum value of the magnetization, but decreases with reduced impurity concentration.

\begin{acknowledgments}
We are grateful to A.~V.~Balatsky and J.~Fransson for discussions. A.B.-S.~was supported by the Swedish Research Council (VR). 
\end{acknowledgments}


\onecolumngrid
\setcounter{page}{-1}
\setcounter{table}{0}
\setcounter{section}{0}
\setcounter{figure}{0}
\setcounter{equation}{0}
\renewcommand{\thepage}{\Roman{page}}
\renewcommand{\thesection}{S\arabic{section}}
\renewcommand{\thetable}{S\arabic{table}}
\renewcommand{\thefigure}{S\arabic{figure}}
\renewcommand{\theequation}{S\arabic{equation}}
\clearpage
\subsection{Supplementary material}

In this supplementary material we give the main steps in the continuum calculation for long-range Coulomb interactions using the coherent potential approximation (CPA) to treat nonmagnetic impurities. We start with the Hamiltonian of a  continuum model including both a finite impurity concentration (disorder) and Coulomb repulsion:
\begin{equation}\label{hamiltonian}
H=\sum\limits_{\bm{k}}\bar{\bm{\psi}}_{\bm{k}}\left(\bm{k}\cdot\bm{\sigma}\right)\bm{\psi}_{\bm{k}}+\dfrac{1}{2}\int d^2\bm{r}_1\int d^2\bm{r}_2\bar{\bm{\psi}}(\bm{r}_1)\bar{\bm{\psi}}(\bm{r}_2)U\left(\vert\bm{r}_1-\bm{r}_2\vert\right)\bm{\psi}(\bm{r}_2)\bm{\psi}(\bm{r}_1)+\int d^2\bm{r}\sum\limits_nV\left(\bm{r}-\bm{r}_n\right)\bar{\bm{\psi}}(\bm{r})\bm{\psi}(\bm{r}).
\end{equation}
The first term in the Hamiltonian gives the linear gapless dispersion relation inherent to Dirac materials with the Pauli matrix $\bm{\sigma}$ operating in spin-space for TIs (in agreement with the main text we put $\hbar v_F=1$). The second term is the electron-electron interaction $U(\vert\bm{r}_1-\bm{r}_2\vert)=e^2/(\epsilon_0 \vert\bm{r}_1-\bm{r}_2\vert)$. The last term describes the effect of impurities, where we only consider short-range diagonal disorder with $V(\bm{r}-\bm{r}_n)=V\delta\left(\bm{r}-\bm{r}_n\right)$. 

The on-site Green's function for the unperturbed Dirac Hamiltonian (first term in Eq.~(\ref{hamiltonian})) is
\begin{equation}\label{green}
g_0(\varepsilon)=\langle\bm{r}\sigma\vert G(\varepsilon)\vert\bm{r}\sigma\rangle=\dfrac{1}{N}\sum\limits_{\bm{k}}\dfrac{\varepsilon}{\varepsilon^2-k^2}=-\dfrac{\varepsilon}{4\pi}\log\left\vert\dfrac{\varepsilon_c^2-\varepsilon^2}{\varepsilon^2}\right\vert-\dfrac{i|\varepsilon|}{4}\theta\left(\varepsilon_c-|\varepsilon|\right),
\end{equation}
where we have used the cut-off parameter $\varepsilon_c$, which can be determined from $\varepsilon_c=2v_F\sqrt{\pi}/a$ for a round-shaped Brillouin zone with the lattice spacing $a$. 
In most cases we are only interested in the asymptotic behavior of Eq.~(\ref{green}) at $|\varepsilon|\ll\varepsilon_c$:
\begin{equation}\label{asympgreen}
g_0(\varepsilon)\approx \dfrac{1}{4\pi v_0^2}\left(2\varepsilon\log\left(\dfrac{|\varepsilon|}{\varepsilon_c}\right)-i\pi|\varepsilon|\right).
\end{equation}

In terms of the impurities, we here presume that the impurities are randomly distributed over the lattice sites, with the on-site potential taking values $V$ and 0 with probabilities $c$ and $1-c$, respectively. We also assume an impurity concentration $0<c\ll 1$, so the results of the calculations can be presented in a power series of $c$. 
For a strongly scattering medium but where the concentration of impurities is low enough we can work within the CPA, which is a single-site approach. The advantage of the CPA is that it includes multiple scattering off the same impurity but that scattering off impurity clusters is totally disregarded. In this approach the bare Green's function (\ref{asympgreen}) is to be replaced by a renormalized Green's function and averaged over impurity configurations. The latter is achieved by defining the self-energy $\sigma_{\rm CPA}(\varepsilon)$, which has restored translational invariance after configurational averaging \cite{Lifshits_sup}. In the problem under consideration the self energy $\sigma_{\rm CPA}(\varepsilon)$ is also site-diagonal and identical for both spins. As a result, the modified propagator averaged over impurities is $\langle g(\varepsilon)\rangle =g_0(\varepsilon-\sigma)$ and the CPA Green's function can thus be written as
\begin{equation}\label{greencpa}
G_{\rm CPA}^\pm \left(\varepsilon,\bm{k}\right)=\dfrac{1}{\varepsilon-\sigma_{\rm CPA}(\varepsilon)-\varepsilon_{\bm k}^\pm},
\end{equation}
where $\varepsilon_{\bm k}^\pm = \pm |{\bm k}|$ is the bare dispersion for the upper and lower Dirac bands. 
In the CPA the on-site potential takes the values $V$ and $0$ depending on whether an impurity is present or not at a lattice site, set by the probabilities $c$ and $1-c$, respectively. Self-consistency is achieved by satisfying $\langle T(\varepsilon-\sigma_{\rm CPA}(\varepsilon))\rangle=0$ for the impurity $T$-matrix, which gives \cite{Skrypnyk2006_sup}
\begin{equation}\label{selfcons}
\sigma_{\rm CPA}(\varepsilon)=\dfrac{cV}{1-\left(V-\sigma_{\rm CPA}(\varepsilon)\right)g_0(\varepsilon-\sigma_{\rm CPA}(\varepsilon))}\approx \dfrac{cV}{1-Vg_0(\varepsilon-\sigma_{\rm CPA}(\varepsilon))},
\end{equation}
where the last identity holds for vacancies. 

As long as the CPA is valid we can evaluate the occupation fraction  $f(\varepsilon_{\bm k}^\alpha)=\int\limits_{-\Lambda}^{\varepsilon_F} \dfrac{d\varepsilon}{\pi}{\rm Im}\, G^\alpha_{\rm CPA}(\varepsilon,\bm{k})$, where $\Lambda$ is a proper regularization and $\alpha = \pm 1$. The ground state energy of Eq.~(\ref{hamiltonian}) can then be calculated (similar to that of Ref.~\cite{Peres2006_sup}) arriving at
\begin{equation}
E=E_{\rm kin}+E_{\rm el-el}=\sum\limits_{\bm{k}\alpha}\varepsilon_{\bm k}^\alpha f(\varepsilon_{\bm k}^\alpha)-\dfrac{\pi e^2}{2S\epsilon_0}\sum\limits_{{\bm k},{\bm p}}\sum\limits_{\alpha_1,\alpha_2}\dfrac{1+\alpha_1\alpha_2\cos\left(\varphi_{\bm{k}}-\varphi_{\bm{p}}\right)}{\vert\bm{k}-\bm{p}\vert}f(\varepsilon_{\bm k}^{\alpha_1})f(\varepsilon_{\bm p}^{\alpha_2}),
\end{equation}
with $\varphi_{\bm k}$ being the angle of ${\bm k}$ relative to the $x$-axis.
From this expression we can calculate the difference in energy between an unpolarized ground state with total occupation $2n$ and that of a state with $n+\delta n$ up-spins and $n-\delta n$ down-spins. When this energy difference is negative the system spontaneously spin-polarizes. For small concentrations of vacancies (${\rm Im}\,\sigma\ll k_c$) we find $g_c \sim 3.7$, to be compared to $g_c \sim 5$ for the clean system.

\end{document}